\documentclass[12pt]{article}
\usepackage[english]{babel}
\usepackage{amsmath,amssymb}
\newtheorem{remark}{Remark}
\newtheorem{example}{Example}
\newtheorem{theorem}{Theorem}
\newtheorem{definition}{Definition}

\usepackage{color}

\begin{document}

\title{On the construction of conservation laws: \\a mixed approach}

\author{M.~Ruggieri$^1$, M.~P.~Speciale$^2$\\
\ \\
{\footnotesize $^1$Faculty of Engineering and Architecture, Kore University of Enna} \\
{\footnotesize Via delle Olimpiadi, Cittadella Universitaria,  I--94100, Enna, Italy}\\
{\footnotesize Email: marianna.ruggieri@unikore.it}\\
{\footnotesize $^2$Department of Mathematical and Computer Sciences,}\\
{\footnotesize Physical Sciences and Earth Sciences, University of Messina}\\
{\footnotesize Viale F. Stagno d'Alcontres 31, I--98166 Messina, Italy}\\
{\footnotesize Email: mpspeciale@unime.it}
}

\date{}
\maketitle

\begin{abstract}
A new approach, combining the Ibragimov method and the one by Anco and Bluman, with the aim of
algorithmically computing local conservation laws of partial differential equations, is discussed.
Some examples of the application of the procedure are given. The method, of course,
is able to recover all the conservation laws found by using the direct method;
at the same time we can characterize which symmetry, if any, is responsible for the existence of
a given conservation law. Some new local conservation laws for the Short Pulse equation and for the Fornberg--Whitham equation are also determined.

\vspace{0.2 cm}
\noindent
{\bf Keywords}: Conservation laws, Lie point symmetries, a systematic procedure.
\end{abstract}


\section{Introduction}
In dealing with differential equations, conservation laws have a deep relevance, since they often express
the conservation of physical quantities. They are also important due to their use in investigating integrability,
existence, uniqueness and stability of solutions, or in implementing efficient numerical methods of integration.
In 1918, Emmy Noether \cite{Noether} presented her celebrated procedure (Noether's theorem) to find local
conservation laws for systems of differential equations arising from a variational principle. Noether
proved that a point symmetry of the action functional (action integral) provides a local conservation law
through an explicit formula that involves the infinitesimals of the point symmetry and the Lagrangian of
the action functional. The most important limitation of Noether's theorem for the determination of local conservation
laws relies on the fact that it applies to variational systems.

In recent years, systematic procedures to find conservation laws also for non variational problems have been
introduced. The first one, the ``direct method'', proposed by Anco and Bluman in 1996 \cite{AncoBluman96} (see also \cite{Anco1997}), gives the possibility to generate local as well nonlocal conservation laws.
A second approach was introduced by Ibragimov in 2007 \cite{Ibragimov2007}, and improved in recent years \cite{IbraAlga2010}-\cite{Ibragimov14} with the formulation of some
theorems allowing for the construction of conservation laws starting from the symmetries of the
differential equations and the use of a {\it formal} Lagrangian. Recently many authors have been employing these concepts in order to establish conservation laws for equations and systems which arises in many fields of the applied sciences \cite{Johnpillai}-\cite{Gandarias3}.

In Noether's theorem, one starts from  the Euler-Lagrange equations
\begin{equation}
\frac{\delta \mathcal{L}}{ \delta u^\alpha}=0,  \quad \alpha =1,\ldots,m,
\label{condL}
\end{equation}
where $\mathcal{L}(\mathbf{x},\mathbf{u},\mathbf{u}_{(1)},\ldots,\mathbf{u}_{(k)})$  is a Lagrangian involving the independent variables $\mathbf{x} =(x_1,\ldots,x_n)$, the dependent variables $\mathbf{ u}=(u^1,\ldots,u^m)$, and the partial derivatives up to a fixed order $k$;
moreover,
$$
\frac{\delta }{\delta u^\alpha}=\frac{\partial }{\partial u^\alpha}+\sum_{j=1}^\infty(-1)^j D_{i_1} D_{i_2}\ldots D_{i_j} \frac{\partial }{\partial u^{\alpha}_{i_1i_2\ldots i_j}}
$$
are the Euler-Lagrange operators; here and the following we use the Einstein convention of sum over repeated indices. Noether's theorem  states that if the variational integral with the Lagrangian $\mathcal{L}$ is invariant under a group $G$  generated by
\begin{equation}
\label{sym}
X=\xi^i \frac{\partial }{\partial x^i}+\eta^{\alpha} \frac{\partial}{\partial u^{\alpha}},
\end{equation}
then the vector field $\mathbf{C}=(C^1,\ldots,C^n)$  defined by
\begin{equation}
\label{Noether}
C^i =  \xi^i \mathcal{ L}+(\eta^\alpha-\xi^i u^\alpha_i) \frac{\partial {\cal L}}{\partial u_i^\alpha}
\end{equation}
provides a conservation law for the Euler-Lagrange equations (\ref{condL}),  \emph{i.e.}, it is
$\mbox{div} \, \mathbf{C} \equiv D_i(C^i)=0$ for all solutions of (\ref{condL}),
\begin{equation}
\left.D_i(C^i)\right|_{(1)} =0,
\label{2.19}
\end{equation}
where $D_i$ is the operator of total differentiation:
\begin{equation}
D_i=\frac{\partial}{\partial x^i}+u^{\alpha}_ i \frac{\partial}{\partial u^{\alpha}_i}+u^{\alpha}_{ ik}\frac{\partial}{\partial u^{\alpha}_{ik}}+\ldots.
\end{equation}
Any vector field $\mathbf{C}$ satisfying (\ref{2.19}) is called a conserved vector.

The plan of the paper is the following. In Section~\ref{sec:direct}, we sketch briefly either the Anco and Bluman direct approach or the Ibragimov method, and discuss their main features;
it is worth of being remarked that the Anco and Bluman method (ABM) often finds more conservation laws that those provided by the Ibragimov method (IM), but the latter approach
is able to establish a direct link between the conservation laws and the symmetries admitted by the differential equations at hand. In Section~\ref{sec:newMethod}, we introduce our approach that in some sense merges these two methods. Our method, in fact, is able to recover, at least in the examples we considered, all the
conservation laws found by using ABM, and at the same time to show which symmetry,
if any, is related to a conservation law. Finally, in Section~\ref{sec:applications}, we provide explicitly some applications of the method.

\section{The Anco--Bluman and Ibragimov methods}
\label{sec:direct}

The ABM \cite{AncoBluman96,Anco1997},\cite{AncoB2002I}-\cite{Anco2010} does not lead directly to the construction of local conservation laws of a given system of differential equations even if the authors suggest several ways of finding the fluxes of local conservation laws.
Given a system of
$\overline{m}$ $k$-th order differential equations
\begin{equation}
F_{\overline{\alpha}}(\mathbf{x}, \mathbf{u}, \mathbf{u}_{(1)},\mathbf{ u}_{(2)},\ldots, \mathbf{u}_{(k)})=0, \qquad {\overline{\alpha}}=1,\ldots, \overline{m},\label{pde0}
\end{equation}
with $m$ dependent variables $\mathbf{u} = (u^1,\ldots,u^m)$, we have the following fundamental result.
\begin{theorem} A set of non-singular local multipliers
\[
v^{\overline{\alpha}}=v^{\overline{\alpha}}(\mathbf{x},\mathbf{u},\mathbf{u}_{(1)},\ldots,\mathbf{u}_{(\ell)})
\qquad \ell \leq k, \;{\overline{\alpha}}=1, \ldots, \overline{m},
\]
yields a divergence expression for the
system (\ref{pde0}) if and only if the set of equations
\begin{equation}
\frac{\delta }{\delta u^\alpha}\left(v^{\overline{\alpha}} F_{\overline{\alpha}}\right)\equiv 0, \qquad
\alpha = 1,\ldots,m, \label{Anco}
\end{equation}
hold for arbitrary function $\mathbf{u}(\mathbf{x})$.
\end{theorem}

First one looks for sets of multipliers of the form
\[
v^{\overline{\alpha}}=v^{\overline{\alpha}}(\mathbf{x},\mathbf{u},\mathbf{u}_{(1)},\ldots,\mathbf{u}_{(\ell)}) \qquad{\overline{\alpha}}=1,\ldots,\overline{m},\; \ell \leq k,
\]
where the dependence of multipliers on their arguments is chosen so that singular multipliers do not arise.
Then  the set of determining equations (\ref{Anco}) for arbitrary $\mathbf{u}(\mathbf{x})$ are solved
to find all such sets of multipliers satisfying the identity
\begin{equation}
v^{\overline{\alpha}} F_{\overline{\alpha}}\equiv D_i \phi^i(\mathbf{x},\mathbf{u},\mathbf{u}_{(1)},\ldots,\mathbf{u}_{(\ell)}).\label{Anco1}
\end{equation}
$\phi^i$ ($i=1,\ldots,n$) being the fluxes of the conservation law.

Once a set of multipliers is known, it is necessary to compute the fluxes of the
local conservation law (\ref{Anco1}). One can achieve this task  by solving the set of determining equations for the fluxes $\phi^i$, or, in the case of complicated forms of multipliers and/or differential equations, through an integral (homotopy) formula.

The second approach \cite{Ibragimov2007} consists in reformulating the Noether's theorem for the determination of conservation laws both in the presence of equations with a variational structure and in the case of differential equations not having variational structure.
\begin{theorem}
If the operator
\begin{equation}
X=\xi^i \frac{\partial }{\partial x^i}+\eta^{\alpha} \frac{\partial}{\partial u^{\alpha}}
\end{equation}
is admitted by the Euler-Lagrange equations
\begin{equation}
\begin{aligned}
\frac{\delta \mathcal{L} }{\delta u^\alpha} =0,\quad \alpha= 1,\ldots,m,
\label{EulLagran}
\end{aligned}
\end{equation}
and satisfies
\begin{equation}
\label{inv}
X(\mathcal{L})+(D_k \xi^k) \mathcal{L}=0,
\end{equation}
then the components
\begin{equation}
\begin{aligned}
C^i &=  \xi^i {\cal L}+W^\alpha\left[\frac{\partial {\cal L}}{\partial u_i^\alpha} - D_j\left(\frac{\partial {\cal L}}{\partial u_{ij}^\alpha} \right)  + D_jD_k\left(\frac{\partial {\cal L}}{\partial u_{ijk}^\alpha} \right)  - \ldots \right]\\
&+ D_j (W^\alpha) \left[\frac{\partial {\cal L}}{\partial u_{ij}^\alpha} - D_k\left(\frac{\partial {\cal L}}{\partial u_{ijk}^\alpha} \right)  + \ldots \right]\\
&+ D_j D_k(W^\alpha)\left[\frac{\partial {\cal L}}{\partial u_{ijk}^\alpha} -  \ldots \right],\qquad i=1,\ldots,n,
\end{aligned}
\label{CONSERVED}
\end{equation}
with
$$ W^\alpha=\eta^{\alpha} - \xi^j u^{\alpha}_j,\qquad \alpha=1,\ldots,m,$$
are the fluxes of the conservation law
$$\left.D_{i}(C^i)\right|_{(\ref{EulLagran})}=0.$$
\end{theorem}
The  proof of the theorem  is  based  on the operator identity
\begin{equation}
X(\mathcal{L})+\mathcal{L}D_i(\xi^i)=W^{\alpha}\frac{\delta \mathcal{L}}{\delta u^{\alpha}}+D_i (\mathcal{N}^i\mathcal{L})
\label{condIb}
\end{equation}
where
$$\mathcal{N}^i =\xi^i +W^\alpha \frac{\delta}{\delta u_i^\alpha}+ \sum_{s=1}^\infty
D_{i_1} \ldots D_{i_s}(W^\alpha) \frac{\delta}{\delta u^\alpha_{i_1 \ldots  i_s}},
\quad  i =1,\ldots,n;$$
taking into account (\ref{EulLagran}), it follows
$$C^i =\mathcal{N}^i(\mathcal{L}), \quad i =1,\ldots,n. $$
Notice that the formal Lagrangian $\mathcal{L}$ needs to be written in symmetric form with respect to all mixed derivatives.

In the case of a first order Lagrangian, (\ref{CONSERVED}) coincides with (\ref{Noether}).
Moreover, one can omit the term $\xi ^i\mathcal{L}$ when it is convenient \cite{Ibragimov14}. This term provides a trivial conserved vector because $\mathcal{L}$ vanishes on the solutions of (\ref{pde0}).

Ibragimov obtains the explicit formula for constructing the conservation laws associated with the symmetries of any nonlinearly self-adjoint system of equations \cite{IbraAlga2010}-\cite{Ibragimov14}.

\begin{definition}
A system of $\overline{m}$ $k$-th order  differential equations
\begin{equation}
F_{\overline{\alpha}}(\mathbf{x}, \mathbf{u}, \mathbf{u}_{(1)}, \mathbf{u}_{(2)}, \ldots, \mathbf{u}_{(k)})=0, \qquad {\overline{\alpha}}=1,\ldots, \overline{m},\label{pde1}
\end{equation}
with $m$ dependent variables $\mathbf{u} = (u^1,\ldots,u^m)$, is said to be nonlinearly
self-adjoint if the adjoint equations
\begin{equation}
\begin{aligned}
&F_\alpha^*(\mathbf{x}, \mathbf{u}, \mathbf{u}_{(1)},  \mathbf{v}_{(1)}, \mathbf{u}_{(2)}, \mathbf{v}_{(2)}, \ldots \mathbf{u}_{(k)}, \mathbf{v}_{(k)})=
\frac{\delta \mathcal{L}}{\delta u^\alpha}=0, \quad \alpha=1,\ldots m,
\end{aligned}
\label{self0}
\end{equation}
($\mathcal{L}=v^{\overline{\beta}} F_{\overline{\beta}}$, with $\overline{\beta}=1, \ldots, \overline{m}$, is the so called {\it formal} Lagrangian), are satisfied for all solutions $\mathbf{u}(\mathbf{x})$ of the original system (\ref{pde1}) upon a substitution
\begin{equation}
v^{\overline{\alpha}} =\psi^{\overline{\alpha}} (\mathbf{x},\mathbf{u}), \quad  {\overline{\alpha}}=1,\ldots,m \label{70}
\end{equation}
such that
$\boldsymbol{\psi}(\mathbf{x},\mathbf{u})\neq0$.
\end{definition}
In other words, the following equations hold true
\begin{equation}
\begin{aligned}
&F_\alpha^\star(\mathbf{x}, \mathbf{u}, \mathbf{u}_{(1)},  \boldsymbol{\psi}_{(1)}, \mathbf{u}_{(2)}, \boldsymbol{\psi}_{(2)}, \ldots \mathbf{u}_{(k)}, \boldsymbol{\psi}_{(k)})\\
&\qquad=\lambda_\alpha^{\overline{\beta}} F_{\overline{\beta}}(\mathbf{x}, \mathbf{u}, \mathbf{u}_{(1)}, \mathbf{u}_{(2)}, \ldots, \mathbf{u}_{(k)}), \qquad \alpha=1,\ldots,m,
\end{aligned}
\label{self}
\end{equation}
or, equivalently, by (\ref{self0}) and (\ref{self}), the equations
\begin{equation}\label{self1}
\frac{\delta( v^{\overline{\beta}}\,F_{\overline{\beta}})}{\delta u^\alpha}=\lambda_\alpha^{\overline{\beta}} F_{\overline{\beta}}(\mathbf{x}, \mathbf{u}, \mathbf{u}_{(1)}, \mathbf{u}_{(2)}, \ldots, \mathbf{u}_{(k)}), \qquad \alpha=1,\ldots,m,
\end{equation}
where $\lambda^{\overline{\beta}}_\alpha$ are undetermined coefficients, and
$$\boldsymbol{\psi}_{(\sigma)} =\{D_{i_1} \ldots D_{i_{\sigma}}(\psi^{\overline{\alpha}} (\mathbf{x},\mathbf{u}))\},\quad \sigma =1,\ldots,k,\quad{\overline{\alpha}}=1,\ldots,\overline{m}. $$

Here, $\mathbf{v} = (v^1,\ldots,v^{\overline{m}})$ and $\boldsymbol{\psi}= (\psi^1,\ldots,\psi^{\overline{m}})$ are  $\overline{m}$-dimensional vectors, and $\boldsymbol{\psi(\mathbf{x},\mathbf{u})}\neq 0$ means that not all components $\psi^{\overline{\alpha}} (\mathbf{x},\mathbf{u})$ (${\overline{\alpha}}=1,\ldots,\overline{m}$) vanish simultaneously.

Ibragimov also considers the case in which the point-wise substitution (\ref{70}) is replaced by
$$v^{\overline{\alpha}} =\psi^{\overline{\alpha}} (\mathbf{x}, \mathbf{u}, \mathbf{u}_{(1)}, \mathbf{u}_{(2)}, \ldots, \mathbf{u}_{(r)}), \quad {\overline{\alpha}}=1,\ldots,\overline{m}, \qquad r \leq k.$$
Then (\ref{self}) will be written, e.g., in the case $r =1$, as follows
\begin{equation}
F_\alpha^*(\mathbf{x}, \mathbf{u}, \mathbf{u}_{(1)}, \boldsymbol{\psi}_{(1)},\mathbf{u}_{(2)}, \boldsymbol{\psi}_{(2)},\ldots, \mathbf{u}_{(k)},\boldsymbol{\psi}_{(k)})=\lambda_\alpha^{\overline{\beta}} F_{\overline{\beta}}+\lambda_\alpha^ {j \overline{\beta}} D_j(F_{\overline{\beta}}),
\label{self2}\end{equation}
where $\lambda^{j\overline{\beta}}_\alpha$ are undetermined coefficients.

Using the definition of nonlinear self-adjointness and the theorem $1$ on conservation laws proved in \cite{Ibragimov2007}, the explicit formula for constructing conservation laws associated with symmetries of any nonlinearly self-adjoint system of equations is obtained.


\begin{theorem}
Let a system of
$\overline{m}$ $k$-th order differential equations
\begin{equation}
 F_{\overline{\alpha}}(\mathbf{x}, \mathbf{u}, \mathbf{u}_{(1)}, \mathbf{u}_{(2)}, \ldots, \mathbf{u}_{(k)})=0, \quad {\overline{\alpha}}=1\dots \overline{m},\label{pde}
\end{equation}
with $m$ dependent variables $\mathbf{u} = (u^1,\ldots,u^m)$, be non linearly self-adjoint.
Specifically, let the adjoint system (\ref{self}) or (\ref{self1}) be satisfied for all solutions of (\ref{pde}) upon
the substitution (\ref{70}).
Then, any Lie point, contact or Lie-B\"acklund symmetry (\ref{sym}),
as well as any nonlocal symmetry of (\ref{pde}), leads to a conservation law  constructed by the formula (\ref{CONSERVED}).
\end{theorem}

\begin{remark}
The Ibragimov method, that allows to write immediately the fluxes $C^i$ ($i=\ldots,n$), often obtains a limited set of conservation laws. The conservation laws obtained are the ones linked to symmetries that are usually point symmetries, since Lie-B\"acklund, generalized and nonlocal symmetries are not easy to determine.
\end{remark}

\subsection{Some applications}
\label{subsec:appl}
Here we report some examples of applications of the two methods for computing local conservation laws.

\begin{example}[KdV equation]
Let us consider the Korteweg-deVries equation
\begin{equation}
\label{kdv}
\Delta\equiv u_t-u_{xxx} - u  u_x=0.
\end{equation}

In \cite{Ibragimov11,Ibragimov14}, it has been proved that KdV equation is nonlinear self-adjoint for
$$\psi=A_1 +A_2u+ A_3(x+tu),$$
where $A_1, A_2, A_3$ are arbitrary constants; hence, the following three linearly independent solutions
$$\psi_1=1, \quad \psi_2=u, \quad\psi_3=(x+tu)$$
are recovered.

Using the symmetries admitted by KdV equation,
\begin{equation}
\begin{aligned}
&\Xi_1=t\frac{\partial}{\partial x}-\frac{\partial}{\partial u}, \quad  \Xi_2=3 t\frac{\partial}{\partial t}+x\frac{\partial}{\partial x}-2 u\frac{\partial}{\partial u}, \quad \Xi_3=\frac{\partial}{\partial t}, \quad \Xi_4=\frac{\partial}{\partial x},
 \end{aligned}\label{KdVsim}
 \end{equation}
some conservation laws are determined. In particular, using the scaling group and $\psi_1$ or $\psi_2$,
\begin{equation}
\label{cl1_kdv}
\begin{aligned}
&D_t\left(u \right) - D_x\left(\frac{1}{2}u_x^2+u_{xx} \right)=0,\\
&D_t\left(u^2 \right) - D_x\left(u_x^2 - 2u\,u_{xx} - {2\over 3}u^3\right)=0,
\end{aligned}
\end{equation}
whereas it is claimed  (see \cite{Ibragimov1985}, Sect.22.5]) that the Galilei group leads to  the  conservation law
\begin{equation}
\label{cl2_kdv}
D_t\left(\frac{1}{ 2}t u^2 +x u\right)-D_x\left(\frac{1}{2}x u^2 +t u u_{xx} - \frac{1}{2}t u^2_x - x u_{xx} +u_x\right) = 0.
\end{equation}

The ABM method \cite{AncoB2002I}-\cite{Anco2010}, assuming $\psi(t,x,u)$, provides the  three multipliers
$$\psi_1=1, \quad \psi_2=u, \quad\psi_3=(x+tu),$$
whereupon, by the relation (\ref{Anco1}),
the conservation laws (\ref{cl1_kdv}) and (\ref{cl2_kdv}) are found \cite{BlumanChe2010}.
\end{example}

\begin{example}

Let us consider the polytropic gas dynamics equations in $n$ space dimensions ($n\leq 3$):
\begin{equation}
\begin{aligned}
&\frac{\partial \rho}{\partial t} +\nabla\cdot(\rho \mathbf{u})=0 ,\\
&\rho\left(\frac{\partial \mathbf{u}}{\partial t}+(\mathbf{u}\cdot\nabla)\mathbf{u}\right)+\nabla p=0,\\
&\frac{\partial p}{\partial t}+\mathbf{u}\cdot\nabla p+\gamma p\nabla\cdot\mathbf{u}=0,
\end{aligned}
\label{2DE}
\end{equation}
where $\rho$ is the mass density, $\mathbf{u}=(u^1,\ldots,u^n)$ the velocity, $p$ the pressure, $\gamma$ the adiabatic index and $\mathbf{x}=(x^1,\ldots,x^n)$ the rectangular space coordinates.


The Lie algebra of point symmetries is spanned by
\begin{equation}
\begin{aligned}
&\Xi_0=\frac{\partial }{\partial t}, \qquad \Xi_i=\frac{\partial }{\partial x^i} \quad i=1,\ldots,n,\\
&\Xi_{4}=t\frac{\partial }{\partial t}+x^k\frac{\partial }{\partial x^k },\\
&\Xi_{5}=t\frac{\partial }{\partial t}-u^k\frac{\partial }{\partial u^k }-2p\frac{\partial }{\partial p},\qquad
\Xi_6=\rho\frac{\partial }{\partial \rho}+p\frac{\partial }{\partial p},\\
&\Xi_{6+i}=t\frac{\partial }{\partial x^i}+\frac{\partial }{\partial u^i }, \quad i=1,\ldots,n,\\
&\Xi_{9+i}=x^k\frac{\partial }{\partial x^j}-x^j\frac{\partial }{\partial x^k}+u^k\frac{\partial }{\partial u^j}-u^j\frac{\partial }{\partial u^k},\qquad j,k=1,\ldots,n, \quad j<k;
\end{aligned}
\label{symgas}
\end{equation}
the
operators $\Xi _{0}$ and $\Xi _{i}$ ($i=1,\ldots,n $) characterize time and space
translations, $\Xi _{4}$, $\Xi _{5}$ and $\Xi _{6}$ scaling
transformations, $\Xi _{6+i}$ ($i=1,\ldots,n$) Galilean
transformations, and the remaining operators characterize spatial
rotations.

In the one-dimensional case, Ibragimov proves that the system is nonlinear self-adjoint with
$$\psi^1 =\rho u,\qquad \psi^2 =
\frac{{u}^2}{ 2}, \qquad \psi^3 =\frac{1}{\gamma-1};$$
the same proof applies also in the case of more than one space dimension.

In a recent paper \cite{AncoIbra15}, the classification of all well known local and nonlocal classical conservation laws, written in the integral form, is listed:
\begin{equation}\label{CLGas}
\begin{aligned}
&\frac{d}{d t}\int_ {\Omega(t)}\rho d \omega =0 , && \hbox{Conservation of mass},\\
&\frac{d}{d t}\int_ {\Omega(t)}\rho \mathbf{u} d \omega= -\frac{d}{d t}\int_ {S(t)}\rho  \nu d S,  &&\hbox{Momentum},\\
&\frac{d}{d t}\int_ {\Omega(t)}\rho (t \mathbf{u} -\mathbf{x}) d \omega= -\int_ {S(t)}t \rho  \boldsymbol{\nu} d S,
&&\hbox{Center of mass},\\
&\frac{d}{d t}\int_{\Omega(t)} (\frac{1}{2}\rho  |\mathbf{u}|^2 +\frac{p}{\gamma-1}) d \omega= -\int_ {S(t)} p  \mathbf{u} \cdot
\boldsymbol{\nu} d S,  &&\hbox{Energy},\\
&\frac{d}{d t}\int_ {\Omega(t)}\rho ( \mathbf{x} \times \mathbf{u}) d \omega =-\int_ {S(t)} p(  \mathbf{x}\times \boldsymbol{\nu}) d S,  &&\hbox{Angular momentum},\\
\end{aligned}
\end{equation}
where
 $\Omega(t)$  is an  arbitrary $n$-dimensional volume, moving with fluid, $S(t)$ is the boundary of the volume $\Omega(t)$ and $\boldsymbol\nu$  is the unit (outer) normal vector to the surface $S(t)$.

 If we write the above conservation laws in the general form
$$\frac{d}{d t} \int_{\Omega(t)}T d \omega =\int_ {S(t)}
(\boldsymbol{\chi} \cdot \boldsymbol\nu)d S, $$
then the differential form of these conservation laws follows, say
$$D_t(T)+\nabla\cdot(\boldsymbol\chi+T \mathbf{u}) =0. $$

Moreover, if $\gamma=\frac{n+2}{n}$, it is proved that the system admits the following two additional conservation
laws:
\begin{equation}\label{CLGasplus}
\begin{aligned}
&\frac{d}{d t}\int_{\Omega(t)}[t (\rho  |\mathbf{u}|^2 +n p)-\rho\mathbf{x}\cdot \mathbf{u}] d \omega= -\int_ {S(t)} p  (2 t\mathbf{u}-\mathbf{x}) \cdot
\boldsymbol{\nu} d S\\
&\frac{d}{d t}\int_{\Omega(t)}[t^2 (\rho  |\mathbf{u}|^2 +n p)-\rho\mathbf{x}\cdot(2 t  \mathbf{u}-\mathbf{x})] d \omega= -\int_ {S(t)} 2 t p  ( t \mathbf{u}-\mathbf{x}) \cdot
\boldsymbol{\nu} d S.
  \end{aligned}
\end{equation}

We notice that if $\gamma=\frac{n+2}{n}$, Euler equations admit an additional Lie point symmetry
\cite{Ovsiannikov}-\cite{MargheritiSpeciale}  spanned by
\begin{equation}\label{pro}
\Xi_{13}=t^2\frac{\partial}{\partial t}+t x^i \frac{\partial}{\partial x^i}-
n t\rho \frac{\partial}{\partial\rho}+(x^i-t u^i)\frac{\partial}{\partial u^i}
-(n+2)t p \frac{\partial}{\partial p}.
\end{equation}
It was shown in \cite{Te75} that these two additional conservation laws together with the previous classical conservation laws provide all pointwise conservation laws of the system, i.e., the conservation laws whose components depend on the variables $t$, $\mathbf{x}$, $\rho$, $\mathbf{u}$,  $p$, and do not involve derivatives of $\rho$, $\mathbf{u}$ and $p$.

\end{example}

\section{A new mixed method}
\label{sec:newMethod}
The new approach we propose starts from Ibragimov method even if it overcomes the concept of nonlinearly self-adjointness.

Considering the Euler-Lagrange equations
\begin{equation}
\frac{\delta\mathcal{L}}{\delta u^\alpha}=0, \qquad \alpha=1,\ldots,m,\label{Lag}
\end{equation}
we work directly with the condition
\begin{equation}
\left.D_{i}(C^{i}+H^i)\right|_{(\ref{Lag})}=0,
\label{conservation}
\end{equation}
where $H^i$ are the components of an additional arbitrary vector field  $\mathbf{H}$ with zero divergence,
and $C^i$ are the components of a conserved vector $\mathbf{C}$ defined in (\ref{CONSERVED}).

In fact, if we add an arbitrary vector field  whose divergence is zero in (\ref{inv}), we have
$$X(\mathcal{L})+ D_k \xi^k \mathcal{L}=-D_i H^i=0;$$
because of the relation (\ref{condIb})
it is
$$W^{\alpha}\frac{\delta \mathcal{L}}{\delta u^{\alpha}}+D_i (\mathcal{N}^i \mathcal{L})=-D_i H^i,$$
whereupon
\begin{equation}D_i(C^i+H^i)=-W^{\alpha}\frac{\delta \mathcal{L}}{\delta u^{\alpha}}\label{newinv};
\end{equation}
as a consequence, we get (\ref{conservation}), where the components
$C^i=\mathcal{N}^i(\mathcal{L})$ ($i=1,\ldots,n$) are given by (\ref{CONSERVED}).

In our approach, when we consider a system of partial differential equations of order $k$,
\begin{equation}
F_{\overline{\alpha}}(\mathbf{x}, \mathbf{u}, \mathbf{u}_{(1)},\mathbf{ u}_{(2)},\ldots, \mathbf{u}_{(k)})=0, \qquad {\overline{\alpha}}=1,\ldots, \overline{m},
\end{equation}
the vector field $\boldsymbol{\psi}$ involved in the expression of the formal Lagrangian $\mathcal{L}$,
\begin{equation}
\mathcal{L}=\sum \psi^{\overline{\alpha}}F_{\overline{\alpha}} ,\label{Lagrangian}
\end{equation}
is an unknown arbitrary function of $\mathbf{x}$, $\mathbf{u}$, and possibly also of the partial derivatives of dependent variables up to a finite order.

In fact, if we start from (\ref{newinv}),
and use \cite{Ibragimov14}
\begin{equation}\label{cond2}
\frac{\delta \mathcal{L}}{\delta u^\alpha}=\lambda^{\overline{\beta}}_{\alpha} F_{\overline{\beta}}+\lambda^{j \overline{\beta}}_{\alpha} D_{j}(F_{\overline{\beta}})+\ldots=0, \quad \alpha=1, \ldots,m,
\end{equation}
we obtain a unique condition that overcomes the condition of nonlinear self-adjointness required as  a prerequisite to determine the components of the conserved vector in Ibragimov approch:
\begin{equation}
\begin{aligned}
&D_{i}(C^i+ H^i)=-W^{\alpha}\left( \lambda^{\overline{\beta}}_{\alpha} F_{\overline{\beta}}+\lambda^{j \overline{\beta}}_{\alpha} D_{j}(F_{\overline{\beta}})+\ldots\right),
\end{aligned}
\end{equation}
that evaluated on the solutions of the system of partial differential equations (\ref{pde}), gives us:
\begin{equation}
\left.D_{i}(C^{i}+H^i)\right|_{F_{\overline{\alpha}}=0}=0\label{ourcond}.
\end{equation}
The above relations is a differential system in $\boldsymbol{\psi}$, $\bold{H}$  and all their
derivatives with respect to the independent and dependent variables.
Forcing to zero the coefficient of all derivatives $u^\alpha_{(1)},u^\alpha_{(2)}, \ldots$, we get a set of differential constraints for $\psi^{\overline{\alpha}}$ and $H^i$ ($\overline{\alpha}=1,\ldots,\overline{m}$, $i=1,\ldots,n$);
by solving this set of differential conditions, we may obtain  the explicit expression  of $\boldsymbol{\psi}$ and $\mathbf{H}$, and simultaneously get the components of the conserved vector.

In this context, the presence of the derivatives of the functions $H^i$
$$\bold{H}_{(\sigma)} =\{D_{i_1} \ldots D_{i_{\sigma}}(H^j (\mathbf{x},\mathbf{u}))\},\quad \sigma =1,\ldots,k,\; j=1,\ldots,n,$$
in the coefficients of $u^\alpha_{(1)},u^\alpha_{(2)},\ldots $ in (\ref{ourcond}) provides  the differential constraints linking
$\psi^{\overline{\alpha}}$ to $H^i$ that obviously are not independent.

Moreover, the components of the new conserved vector $$\mathbf{T}=\mathbf{C}+\mathbf{H}$$ contain the components $C^i$ ($i=1,\ldots,n$), linked to the symmetries, and  $H^i$ that may be independent of the symmetries;
in this way, we recover the components of the conserved vector of Ibragimov and  an additional term that, sometimes is independent of the symmetries.

\section{Applications}
\label{sec:applications}
In this Section, we present some applications of the method outlined in the previous section.

\begin{example}[KdV equation]
Let us consider  the KdV equation (\ref{kdv})
and the formal Lagrangian $\mathcal{L}=\psi(u_t-u_{xxx}-u u_x)$,
where $\psi=\psi(t,x,u)$.

Let us take a linear combination of the admitted Lie point symmetries (\ref{KdVsim}), say
$$\Xi=a_1 \Xi_1+a_2 \Xi_2+a_3 \Xi_3+a_4 \Xi_4,$$
$a_1$, $a_2$, $a_3$ and $a_4$ being arbitrary constants.
Now, let us write $T^1$ and $T^2$ by using the expressions of $C^1$ and $C^2$ given by (\ref{CONSERVED})
\begin{equation}
\begin{aligned}
T^1&=C^1+H^1=W \frac{\partial \mathcal{L}}{\partial u_t}+H^1, \\
T^2&=C^2+H^2=\\
&=W \left[\frac{\partial \mathcal{L}}{\partial u_x} - D_{xx}\left(\frac{\partial
\mathcal{L}}{\partial u_{xxx}}\right)\right] -
  D_x W \left[D_x\left(\frac{\partial \mathcal{L}}{\partial u_{xxx}}\right) \right] \\
& \quad+ D_{xx} W  \left[\frac{\partial \mathcal{L}}{\partial u_{xxx}}\right]+H^2,
\end{aligned}
\end{equation}
where $H^1=H^1(t,x,u)$, $H^2=H^2(t,x,u)$,  and
$$W=-(a_1+2 a_2 u)-(a_4+a_1 t+a_2 x) u_x-(a_3+3 a_2 t) u_t.$$

By requiring
$$[D_t(T^1)+ D_x(T^2)]|_{\Delta=0}=0,$$
and forcing to zero the coefficients of the derivatives of $u$, we get the differential constraints for the functions $\psi$, $H^1$ and $H^2$.

If $a_3+3 a_2 t\neq0$ we obtain
\begin{equation}
\begin{aligned}
\psi&=\frac{c_0}{a_2}+ c_2 u + c_3 (t u + x)+g(\xi)(a_3+3 a_2 t)^{1/3}\\
&-c_1\left(\frac{3 a_1 a_3 - a_2(6 a_4 - t (a_1+ 2 a_2 u))}{6 a^2_2}\right)\\
&- c_1\left(\frac{a_1 a_3 - a_2 (3 a_4 + a_3 u + 3 a_2 (t u + x))}{9 a_2^2}\right) \log|a_3 + 3 a_2 t|,\\
H^1&=\left((c_1 \xi+a_2 g(\xi)-g^{\prime \prime \prime}(\xi)) (a_3 + 3 a_2 t)^{1/3}\right.\\
&\left. - \left(a_4 + a_1 t +a_2 x+ \frac{a_3 + 3 a_2 t}{2} u\right)g^{\prime}(\xi)\right)\,u\\
& +a_1(a_3 + 3 a_2 t)^{1/3}g(\xi) +a_1\left(\frac{c_0}{a_2}-c_3 x\right)+\partial_x h_1(t,x)\\
&-c_1\frac{a_1}{3 a_2^3}\left(\frac{a_1}{2} (3 a_3 + a_2 t) - 2 a_2 a_4\right.\\
&\left.+\left(\frac{a_1 a_3}{3} - a_2 (a_4 + a_2 x)\right)
\log(a_3 + 3 a_2 t)\right),\\
H^2&=\left((c_1\xi+a_2 g(\xi)- g^{\prime\prime\prime}(\xi) )(a_3+3 a_2 t)^{1/ 3}\right.\\
&\left.-\left(a_4 + a_1 t +a_2 x+ \frac{a_3 + 3 a_2 t}{2} u\right )g^{\prime}(\xi)\right)
\times\frac{a_4 + a_1 t + a_2 x}{a_3 + 3 a_2 t}u\\
&-\frac{a_1}{(a_3+3 a_2 t)^{1/3}}g^{\prime\prime}(\xi)+h_2(t) -\partial_t h_1(t,x),
\end{aligned}
\end{equation}
where $$\xi=\frac{-a_1 (a_3 + a_2 t) + 2 a_2 (a_4 + a_2 x)}{2 a_2^2 (a_3 + 3 a_2 t)^{
 1/3}},$$  while $g(\xi)$, $h_1(t,x)$ and $h_2(t)$ are arbitrary functions of their arguments and $c_i$ for $i=0,\ldots,3$, arbitrary  constants.
\par
We remark that if
$$c_1=g(\xi)=0,\quad   h_1(t,x)=a_1\left(c_3 \frac{x^2}{2}-\frac{c_0}{a_2}\right)x+h_3(t),$$
and
$$h_2(t)=h^{\prime}_3(t),$$
we obtain
$$H^1=H^2=0, $$
($h_3(t)$ arbitrary function), and we recover exactly Ibragimov expression of $\psi$.

The density and the flux, without the additional terms that give trivial contributions to the conservation laws,
read
\begin{equation}
\begin{aligned}
T^1&=u\left(-c_0+c_1\left(x+t \frac{u}{2}\right)-c_2 \left(a_1+\frac{3}{2} a_2 u\right)+c_3\left(a_4+a_3 \frac{u}{2}\right)\right),\\
T^2&=c_0\left( \frac{1}{2}u^2+u_{xx}\right)\\
&+c_1\left(- \left(\frac{1}{2}x+\frac{1}{3} t u\right)u^2+\left(1+ \frac{1}{2}u_x\right)u_x- \left(x+ t u\right)u_{xx}\right)\\
&+c_2\left(\left(\frac{a_1}{2} +a_2 u\right)u^2-\frac{3}{2} a_2 u_x^2+ (a_1 +3 a_2 u) u_{xx}\right)\\
&+c_3\left(-\left(\frac{1}{2} a_4+\frac{2}{3} a_3 u\right) u^2+\frac{1}{2} a_3 u_x^2- (a_4 + a_3 u) u_{xx}\right)+h_2(t).
\end{aligned}
\end{equation}
By taking $a_1c_2=1$ and all the remaining coefficients vanishing (or $a_4c_3=1$, and all the remaining coefficients vanishing), we get
\begin{equation}
T^1=-u, \qquad T^2=\left(\frac{u^2}{2}+u_{xx}\right)
\end{equation}
related to galilean group (space translation, respectively). Moreover, by taking $a_2c_2=1$ and all the remaining coefficients vanishing (or $a_3c_3=1$, and all the remaining coefficients vanishing), we get
\begin{equation}
T^1=-\frac{3}{2}u^2,
\qquad T^2= -\frac{3}{2} \left(u_x^2-2 u u_{xx}-\frac{2 u^3}{3}\right),
\end{equation}
related to the scaling group (the time translation, respectively).

If only $c_0\neq 0$, we recover the same conservation law linked to the galilean group, whereas if only $c_1\neq 0$ we find the conserved vector
$$T^1=\left(x+t\frac{u}{2}\right)u,
\qquad T^2=\left(u_x+t\left(\frac{u_x^2}{2} -u u_{xx}-\frac{u^3}{3}\right)-x\left(\frac{u^2}{2}+u_{xx}\right)\right)$$
not related to a Lie symmetry but obtained in \cite{Anco1997,Anco2010}.

\end{example}

\begin{example}[The Fornberg--Whitham equation]

Let us consider the Fornberg --Whitham equation
\begin{equation}\Delta=u_t - u_{txx} - u u_{xxx} - 3 u_x u_{xx} + u u_x + u_x = 0\label{FWeq}.
\end{equation}
In a recent paper  \cite{IbragimovValenti}, using the IM,  it was  proved that (\ref{FWeq}) is nonlinear self-adjoint with $\psi=k_0$ ($k_0$ constant), even if only trivial conservation laws have been obtained \cite{Hashemi14}. Now we show that the method proposed in the previous section is able to provide
nontrivial conservation laws for the Fornberg-Whitham equation (\ref{FWeq}).

The equation (\ref{FWeq}) admits the following symmetries:
$$ \Xi_1=\frac{\partial}{\partial t}, \qquad \Xi_2=\frac{\partial}{\partial x}, \qquad \Xi_3=t\frac{\partial}{\partial x}+\frac{\partial}{\partial u}.$$
Starting from a linear combination of them, $\Xi=a_1\Xi_1+a_2 \Xi_2+a_3\Xi_3$ ($a_1$, $a_2$, $a_3$ arbitrary constants), we have
$$W=a_3-a_1u_t-(a_2+a_3 t) u_x.$$
Assuming $\psi$, $H^1$ and $H^2$ functions of $t,x,u$,
through the condition
$$\left.[D_{t}(T^1)+D_x(T^2)]\right|_ {\Delta=0}=0,$$ where
we force to zero the coefficients of all derivatives of $u$,  we obtain
\begin{equation}
\begin{aligned}
&\psi =c_0 t + c_1 x + g(x),\\
&H^1=a_1 (u (c_0+c_1+g^{\prime}(x))+\frac{u^2}{2} (c_1+ g^{\prime}(x)-g^{\prime\prime\prime}(x))), \\
& H^2=0,
\end{aligned}
\end{equation}
with $a_2=a_3=0$ and $a_1\neq0$ (it means that the galilean group and space translation do not induce  local conservation laws), whereas
$g(x)$ is an arbitrary function of its arguments while $c_0$ and $c_1$ are arbitrary constants.

Eliminating the terms that lead to trivial contributions in the conservation law, we get
\begin{equation}
\begin{aligned}
&T_1=a_1\left(c_0 u+\frac{5}{3}c_1 u\left(1+\frac{u}{2}\right)-\frac{5}{3}(c_0 t+c_1 x)u_t\right),\\
&T_2=-\frac{2}{3}c_0 a_1\left(u\left(1+\frac{1}{2}u\right)-u_{x}^2-u_{tx}\right)+\frac{5}{3} c_1 a_1(u_t u_x+u_{xx}+u u_{tx})\\
&\quad-\frac{5}{3}(c_0 t+c_1 x)a_1 \left(u_t(1+u-u_{xx})-2 u_x u_{tx}-u_{ttx}-u u_{txx}\right).
\end{aligned}
\end{equation}
By taking $a_1c_0=1$ and all the remaining coefficients vanishing, we obtain
\begin{equation}
\begin{aligned}
&T^1=u-\frac{5}{3}t u_t, \\ &T^2=-\frac{2}{3}\left(u\left(1+\frac{u}{2}-u_{xx}\right)-u_x^2-u_{tx}\right)\\
&\qquad-\frac{5}{3}t(u_t(1+u-u_{xx})-2 u_x u_{tx}-u u_{txx}-u_{ttx}).
\end{aligned}
\end{equation}
While
if $a_1c_1=1$ and all the remaining coefficients vanishing, we  get
\begin{equation}
\begin{aligned}
&T^1=\frac{5}{3}(u\left(1+\frac{u}{2}\right)-x u_t), \\
&T^2=\frac{5}{3} (u_t u_x+u_{xx}+u u_{tx})-\frac{5}{3}x\left(u_t(1+u-u_{xx})-2 u_x u_{tx}-u u_{txx}-u_{ttx}\right).
\end{aligned}
\end{equation}

We observe that, when
$$g(x)=k_0=const., \qquad c_0=c_1=0,$$ we obtain $H^1=H^2=0$, and the trivial conservation law reported in \cite{Hashemi14} arises.
\end{example}

\begin{example}[Short Pulse equation]

Let us consider Short Pulse equation \cite{Short1}-\cite{Short3}:
\begin{equation}
\Delta\equiv u_{tx}-u-\frac{1}{2}u^2 u_{xx}-u u_x^2=0,\label{short}
\end{equation}
which describes the propagation of linearly polarized ultra-short light pulses in a one-dimensional medium
with assuming that the light propagates in the infrared range.

The Lie point symmetries admitted by (\ref{short}) are spanned by:
$$\Xi_1=\frac{\partial}{\partial t}, \qquad \Xi_2=\frac{\partial}{\partial x}, \qquad \Xi_3=t\frac{\partial}{\partial t}-x\frac{\partial }{\partial x}-u\frac{\partial}{\partial u}.$$

In this case, we consider only the operator $\Xi_3$ so that $$W=-a_3(u+tu_t-xu_x);$$ we neglect time and space translations (operators $\Xi_1$ and $\Xi_2$), since we can insert their contributions at the end of the procedure by replacing $a_3t$ with $a_3t+a_1$ and $a_3x$ with $a_3x+a_2$.

It is well known that the ABM and IM, when $\psi$ depends on the independent, the dependent variables and the first derivatives, lead to obtain some nontrivial conservation laws \cite{Ibragimov14}; here, we consider $\psi=\psi(t,x,u,u_t,u_x)$, $H^1=H^1(t,x,u,u_t,u_x)$ and $H^2=H^2(t,x,u,u_t,u_x)$.

According to the procedure, we impose the constraint
$$\left.[ D_t (T^1)+D_x (T^2)]\right|_{\Delta=0}=0,$$
and solving the differential conditions obtained by requiring that the coefficients of all derivatives of $u$ greater than or equal to $2$ are zero, we get
\begin{equation}
\begin{aligned}
&\psi=\frac{(c_2+3 a_3 f(t))u}{a_3 t}
+(2u_t-u^2 u_x)f(t)+ \frac{c_3 u_x}{\sqrt{1+ux^2}}-u f^\prime(t),\\
&H^1=c_1 \sqrt{1+u_x^2}
+(c_2+a_3 (3f(t)- t f^\prime(t)) )\left( \frac{u+x u_x}{t}-\frac{u^2}{2} u_x\right)u_x\\
&\quad-a_3 u\left((t u +2u_x) f^\prime(t) -t u_x f^{\prime\prime}(t)\right),\\
&H^2=-\frac{u^2}{2}\left(c_1 \sqrt{1+u_x^2}
+(c_2+ a_3 (3 f(t)- t f^\prime(t)) )u_x\left( \frac{u+x u_x}{t}-\frac{u^2}{2} u_x\right) \right.\\
&\quad-a_3 \left(u(t u +2u_x) f^\prime(t) -t u u_x f^{\prime\prime}(t)+ \frac{f^\prime(t)}{t}(2+t^2\frac{u^2}{2})-3 f^{\prime\prime}(t)+t f^{\prime\prime\prime}(t)\right)\\
&\quad \left.+2(c_2+a_3 (3 f(t)- 2t f^\prime(t)))\left(\frac{1}{t^2}+\frac{u^2}{4}-\frac{x}{t} \right)\right),
\end{aligned}
\end{equation}
with $c_1$, $c_2$ and $c_3$ arbitrary constants, and $f(t)$ arbitrary function of $t$.

 When $c_1=c_3=0$, $f(t)=c_0$ ($c_0$ constant) and $c_2=-3 a_3 c_0$, we get $H^1=H^2=0$, and we recover the expression of $\psi$ for which equation (\ref{short}) is nonlinear self-adjoint \cite{IbraAlga2010}-\cite{Ibragimov14}; this is  also the form of the multiplier that leads to get a set of conservation laws by using ABM.

Finally, neglecting the terms leading to trivial conservation laws, we obtain
\begin{equation}
\begin{aligned}
&T^1=c_1 \sqrt{1+u_x^2}-c_2 u^2
+a_3 c_3\frac{(u+t u_t-x u_x)}{(1+u_x^2) \sqrt{1+u_x^2}}u_{xx}+ 2 a_3 t f(t) u u_t,\\
&T^2=-c_1\frac{u^2}{2}\sqrt{1+u_x^2}+ \frac{c_2-a_3 (f(t)+ t f^\prime(t))}{4}(u^4+(u^2 u_x-2 u_t)^2)\\
&\quad-a_3t f(t) \left(u^3 u_t+ (u^2 u_x-2 u_t)(u u_x u_t-u_{tt}+\frac{u^2}{2} u_{tx}))\right)-\frac{a_3 c_3}{\sqrt{1+u_x^2}}\\
&\quad\times \left(\frac{(u+t u_t-x u_x) u_{tx}}{1+u_x^2}-
 (u^2+t u u_t)(1+u_x^2)+ u_x( 2 u_t+t u_{tt}-t \frac{u^2}{2} u_{tx})\right);
\end{aligned}
\end{equation}
moreover, we can replace $a_3t\rightarrow a_3t+a_1$ and $a_3x \rightarrow a_3x+a_2$ and obtain richer forms.

The expressions of density and flux include the known results obtained by applying IM and ABM that we recover if
$c_1=c_3=0$, $f(t)=c_0$ ($c_0$ constant) and $c_2=-3 a_3 c_0$; in fact, with these assumptions
we get the following  conservation law
$$D_t (u^2)+\frac{1}{4}D_x(u^4+(u^2 u_x-2 u_t)^2)=0.$$

Moreover, we get a new conservation law by taking $c_1\neq 0$  and all remaining coefficients vanishing,
$$\left.[D_t(\sqrt{1+u_x^2})-D_x (\frac{u^2}{2}\sqrt{1+u_x^2})]\right|_{\Delta=0}=0.$$
Another conservation law, linked to the scaling group $\Xi_3$, is characterized by the
following expressions of density and flux
\begin{equation}
\begin{aligned}
&T^1= c_3\frac{(u+t u_t-x u_x)}{(1+u_x^2) \sqrt{1+u_x^2}}u_{xx}+ 2 t f(t) u u_t\\
&T^2= \frac{- (f(t)+ t f^\prime(t))}{4}(u^4+(u^2 u_x-2 u_t)^2)\\
&\quad-t f(t) \left(u^3 u_t+ (u^2 u_x-2 u_t)(u u_x u_t-u_{tt}+\frac{u^2}{2} u_{tx}))\right)-\frac{ c_3}{\sqrt{1+u_x^2}}\\
&\quad\times \left(\frac{(u+t u_t-x u_x) u_{tx}}{1+u_x^2}-
 (u^2+t u u_t)(1+u_x^2)+ u_x( 2 u_t+t u_{tt}-t \frac{u^2}{2} u_{tx})\right).
\end{aligned}
\end{equation}
\end{example}

\begin{example}[Euler equations of gas dynamics]
Let us again consider the polytropic gas dynamics equations (\ref{2DE}) with $\gamma=\frac{2+n}{n}$,
whose symmetries  are listed in section \ref{subsec:appl} (formulas (\ref{symgas}), (\ref{pro})).
As usually, we consider a linear combination of all operators admitted by the system, $$\Xi=\sum_{k=0}^{10+i}a_k \Xi_k,\quad i=1,\ldots,n.$$

Following the method above proposed, assuming $\psi^k$, $k=1$,$\ldots$, $2+n$, and $H^i$,  $i=1,\ldots,1+n$, functions of all dependent and independent variables, imposing the condition
$$\left.[D_t(T^1)+D_{x^k}(T^k)]\right|_{(\ref{2DE})}=0, \quad k=1,\ldots,n,$$
and solving the differential constraints, obtained requiring the coefficients of all derivatives to be zero, we get explicit forms of $H^i$ ($i=1,\ldots,1+n$), and, consequently, some conservation laws.

In the following, we consider the $3$-dimensional case that includes the subcases $2$- and $1$-dimensional ones if $x^3=u^3=0$ and all dependent variables functions of $t,x^1,x^2$, or $x^2=x^3=u^2=u^3=0$ with dependent variables function of $t,x^1$, respectively; so, we get:
\begin{equation}
\begin{aligned}
&H^1=-a_6 (\psi^5 p + \psi^1 \rho)+((2 k_6 t+k_5)t+k_3)(\frac{2}{\gamma-1}p +\rho (u^2+v^2+w^2))\\
&\qquad+\rho t(k_7 u + k_{10} v + k_{12} w -
   2 k_6 (u x + v y + w z))\\
   &\qquad+\rho \left(k_{9} u + k_{11} v + k_{8} w-k_5 (u x+vy+w z)\right.\\
&\qquad- k_2 (w x +u z)- k_1 (w y -v z) -k_4 (uy-v x) \\
&\qquad \left. +k_6 (x^2+y^2+z^2)-k_7 x - k_{10}  y + k_{12} z +f(p \rho^{-\gamma}) \right),\\
&H^2=u H^1-a_6 \psi^2 p
+p\left(t(k_7 - 2 k_6 x + k_4 y)+ k_{9} - k_5 x  + k_2 z\right),\\
&H^3=v H^1-a_6 \psi^3 p
+p \left(t(k_{10}  - k_4 x - 2 k_6 y)+ k_{11} - k_5 y + k_1 z\right),\\
&H^4=w H^1-a_6 \psi^4 p
+p\left(t(k_{12} - 2 k_6 z) + k_{8} - k_2 x - k_1 y - k_5 z\right),
\end{aligned}
\end{equation}
with $a_i=0$ for all $i=1,\ldots,13$ and $i\neq6$; for the sake of simplicity, we renamed the variables
as follows: $x^1=x$, $x^2=y$, $x^3=z$, $u^1=u$, $u^2=v$ and $u^3=w$.

In previous formulas, all $\psi^i$ ($i=1,\ldots,5$) continue to be arbitrary functions of their arguments, $f(p \rho^{-\gamma})$ is an arbitrary function of its arguments, and $k_j$ for $j=1,\ldots,12$ are arbitrary constants.

As a result, we have
\begin{equation}
\begin{aligned}
&T^1=((2k_6 t+k_5)t+k_3)(\frac{2}{\gamma-1}p +\rho (u^2+v^2+w^2))\\
&\qquad+\rho t(k_7 u + k_{10} v + k_{12} w -
   2 k_6 (u x + v y + w z))\\
&\qquad+\rho \left(k_{9} u + k_{11} v + k_{8} w-k_5 (u x+vy+w z)\right.\\
&\qquad- k_2 (w x +u z)- k_1 (w y -v z) -k_4 (uy-v x) \\
&\qquad \left. +k_6 (x^2+y^2+z^2)-k_7 x - k_{10}  y + k_{12} z +f(p \rho^{-\gamma}) \right),\\
&T^2=
u T^1+p\left(t(k_7 - 2 k_6 x + k_4 y)+ k_{9} - k_5 x  + k_2 z\right),\\
&T^3=
v T^1+p\left(t(k_{10}  - k_4 x - 2 k_6 y)+ k_{11}  - k_5 y + k_1 z)\right),\\
&T^4=
wT^1+p\left(t(k_{12} - 2 k_6 z) + k_{8} - k_2 x - k_1 y - k_5 z\right),
\end{aligned}\label{gasconserv}
\end{equation}
where we neglected the terms  giving trivial contributions to $T^1$, $T^2$, $T^3$ and  $T^4$.


Splitting the coefficients of the integration constants $k_1,\ldots,k_{12} $, we get twelve forms of the  conserved vectors.

Integrating the density and the fluxes of (\ref{gasconserv}) on $\Omega(t)$, an arbitrary $3$--dimensional volume, moving with the fluid, we obtain all well known classical conservation laws listed in \cite{AncoIbra15}, and reported in section \ref{subsec:appl} (formulas (\ref{CLGas}) and (\ref{CLGasplus})); when all but one coefficient $k_i$ ($i=1,\ldots,12$) and $f(p \rho^{-\gamma})$ vanish, we  get
the conservation of the components of angular momentum ($k_1\neq 0$, $k_2\neq0$, or $k_4\neq0$),   the conservation of energy ($k_3\neq 0$), the  two ``additional'' laws ($k_5\neq 0$, or $k_6\neq 0$),
the laws of center of mass ($k_7\neq 0$, $k_{10}\neq 0$, or $k_{12}\neq 0$),  the conservation of the components of linear momentum ($k_{9}\neq 0$, $k_{11}\neq 0$, or $k_{8}\neq 0$), and  we get the conservation of mass (all $k_i$ vanishing and $f(p \rho^{-\gamma})=const.$).
In addition, a new conservation law arises if $f(p \rho^{-\gamma})$ is not constant.

\end{example}

\section{Conclusions}
In this paper, we introduced a new mixed method for the construction of conservation laws of differential equations. The technique, in  some sense, merges the well known Ibragimov method and the one by Anco and Bluman; our method, in fact, is able to recover, at least in the examples we considered, all the conservation laws found by using "the direct method", and at the same time to show which symmetry, if any, is related to a conservation law. In particular, in section \ref{sec:applications}, we have found with our method new explicit conservation laws for the Short Pulse equation and for the Fornberg--Whitham equation.

\section*{Acknowledgments}
The authors acknowledge the financial support by G.N.F.M. of I.N.d.A.M. through the project "Formazione di pattern, insorgenza di fenomeni oscillatori e soluzioni localizzate in sistemi reazione-diffusione con diffusione non lineare", 2015.


\begin{thebibliography}{99}

\bibitem{Noether} E.~Noether, Invariante Variationsprobleme. Nachr. v. d. Ges. d. Wiss. zu G\"ottingen, Math. Phys.
Kl. 1918, 235--257; English translation, Transp. Th. Stat. Phys., \textbf{1}, 186--207 (1997).


\bibitem{AncoBluman96} S.~C.~Anco, G.~Bluman, Derivation of conservation laws from nonlocal symmetries of differential equations, Journal of Mathematical Physics 37, 2361 (1996); doi: 10.1063/1.531515.

\bibitem{Anco1997} S.~Anco, G.~Bluman, Direct construction of conservation laws from field equations, Phys. Rev. Lett. 78, 2869--2873 (1997).

\bibitem{Ibragimov2007} N.~H.~Ibragimov, A new conservation theorem, J. Math. Anal. Appl. 333, 311--328 (2007).

\bibitem{IbraAlga2010} N.~H.~Ibragimov, Nonlinear self-adjointness in constructing conservation laws, Arch. ALGA 7(8), 1--99 (2010--2011).

\bibitem{Ibragimov11} N.~H.~Ibragimov, Nonlinear self-adjointness and conservation laws, J. Phys. A: Math. Theor. 44, 432002 (2011).

\bibitem{Ibragimov14} N.~H.~Ibragimov, Construction of Conservation Laws Using Symmetries, Springer (2014).

\bibitem{Johnpillai} A.~G.~Johnpillai, C.~M.~Khalique, Symmetry Reductions, Exact Solutions, and Conservation Laws of a Modified Hunter-Saxton Equation, Vol. 2013, Article ID 204746, (2013).

\bibitem{Gandarias}  M.~L.~Gandarias, C.~M.~ Khalique, Nonlinearly Self-Adjoint, Conservation Laws and Solutions for a Forced BBM Equation,  Vol. 2014, Article ID 630282, (2014).

\bibitem{ConservationRS} M.~Ruggieri, M.~P.~Speciale, Conservation laws for a model derived from two layer-fluids, Journal of Physics: Conference Series 482, 012037, 10 (2014).

\bibitem{Gandarias2} M.~Rosa, M.~S.~Bruzon, M.~L.~Gandarias, Lie Symmetry Analysis and Conservation Laws for a Fisher Equation with Variable Coefficients, Applied Mathematics and Information Sciences 9(6), 2783-2792 (2015).

\bibitem{Gandarias3} M. ~L. ~Gandarias, M. ~S. ~Bruzon, and M. ~Rosa  Symmetries and Conservation Laws for Some Compacton Equation, Mathematical Problems in Engineering
Vol. 2015, Article ID 430823, 6 pages  (2015)

\bibitem{AncoB2002I} S.~C.~Anco, G.~Bluman, Direct construction method for conservation laws of partial differential equations I: Examples of conservation law classifications,  Euro. J. Appl. Math. 13, 545-566 (2002).

\bibitem{AncoB2002II} S.~C.~Anco, G.~Bluman, Direct construction method for conservation laws of partial differential equations II: General treatment, Euro. J. Appl. Math. 13, 567-585 (2002).

\bibitem{Anco2010} G.~W.~Bluman, A.~F.~Cheviakov, S.~C.~Anco, Construction of conservation laws: how the direct method generalizes Noether’s theorem. 4th Workshop Group Analysis of Diﬀerential Equations \& Integrable Systems, 13--35 (2009).



\bibitem{Ibragimov1985} N.~Ibragimov, Transformation Groups in Mathematical Physics. Nauka, Moscow(1983) (English trans.:Transformation Groups Applied to Mathematical Physics. Reidel, Dordrecht (1985)).


\bibitem{BlumanChe2010} G.~Bluman, A.~Cheviakov, S.~Anco, Applications of Symmetry Methods to Partial Differential Equations. Springer, New York (2010).


\bibitem{AncoIbra15} S.~Anco, N.~H.~Ibragimov, K.~V.~Imamutdinovab, E.~N.~Karimovab, Solutions of gasdynamic equations associated with classical and new conservation laws, Applied Mathematics and Computation 268, 52-58 (2015).


\bibitem{Ovsiannikov}  L.~V.~Ovsiannikov, Group analysis of differential equations, Academic Press, New York (1982).

\bibitem{OliveriSpeciale2002} F..~ Oliveri, M..~P..~Speciale, Exact solutions to the unsteady equations of perfect gases through Lie group analysis and substitution principles, International Journal of Non-Linear Mechanics 37,257--274 (2002).



\bibitem{MargheritiSpeciale} L.~ Margheriti, M.~P.~Speciale, Unsteady Solutions of Euler Equations Generated by Steady Solutions, Acta Appl. Math. 113, 289–-303 (2011).

\bibitem{Te75} E.~Terentyev, J.~Shmyglevskii,  A complete system of equations in divergence form for the dynamics of an ideal gas. Zh. Vychisl. Mat. i Mat. Fiz. 15 1535–1544 (1975) (English transl., USSR Comput. Math. and Math. Phys. 15 (1975) 167–176).

\bibitem{IbragimovValenti} N.~H.~Ibragimov, R.~S.~Khamitova, A.~Valenti, Self-adjointness of a generalized Camassa--Holm equation, Applied Mathematics and Computation, Vol. 218, Issue 6, 2579–-2583 (2011).

\bibitem{Hashemi14} M.~S.~Hashemi, A.~Haji-Badali, P.~Vafadar, Group Invariant Solutions and Conservation Laws of the Fornberg-Whitham Equation. Z. Naturforsch. A, 69, 489--496 (2014).

\bibitem{Short1} T.~Schäfer, C.~Wayne, Propagation of ultra-short optical pulses in cubic nonlinear media. Phys. D 196, 90–-105 (2004).

\bibitem{Short2} A.~Sakovich, S.~Sakovich, Solitary wave solutions of the short pulse equation. J. Phys A: Math. Gen. 39, L 361–-367 (2006).

\bibitem{Short3} A.~Sakovich, S.~Sakovich, The short pulse equation is integrable. J. Phys. Soc. Jpn. 74, 239 –-241 (2005).

\end{thebibliography}
\end{document}